# PyMOLfold: Interactive Protein and Ligand Structure Prediction in PyMOL

**Colby T. Ford**[1,2,3,4,✉], **Samee Ullah**[5], **Dinler Amaral Antunes**[6], and **Tarsis Gesteira Ferreira**[7]

[1]University of North Carolina at Charlotte, Center for Computational Intelligence to Predict Health and Environmental Risks (CIPHER), Charlotte, NC, USA
[2]University of North Carolina at Charlotte, Department of Bioinformatics and Genomics, Charlotte, NC, USA
[3]University of North Carolina at Charlotte, School of Data Science, Charlotte, NC, USA
[4]Tuple LLC, Charlotte, NC, USA
[5]National Center for Bioinformatics (NCB), Islamabad 45320, Pakistan
[6]University of Houston, Departments of Biology and Biochemistry, Houston, 77204, TX, USA
[7]College of Optometry, University of Houston, 4401 Martin Luther King Boulevard, Houston, TX, 77204-2020, USA

## Abstract

**PyMOLfold is a flexible and open-source plugin designed to seamlessly integrate AI-based protein structure prediction and visualization within the widely used PyMOL molecular graphics system. By leveraging state-of-the-art protein folding models such as ESM3, Boltz-1, and Chai-1, PyMOLfold allows researchers to directly predict protein tertiary structures from amino acid sequences without requiring external tools or complex workflows. Furthermore, with certain models, users can provide a SMILES string of a ligand and have the small molecule placed in the protein structure. This unique capability bridges the gap between computational folding and structural visualization, enabling users to input a primary sequence, perform a folding prediction, and immediately explore the resulting 3D structure within the same intuitive platform.**

**Protein Structure Prediction | PyMOL Plugin
Biomolecular Interaction Modeling | Structure Prediction
Correspondence:** *colby.ford@charlotte.edu*

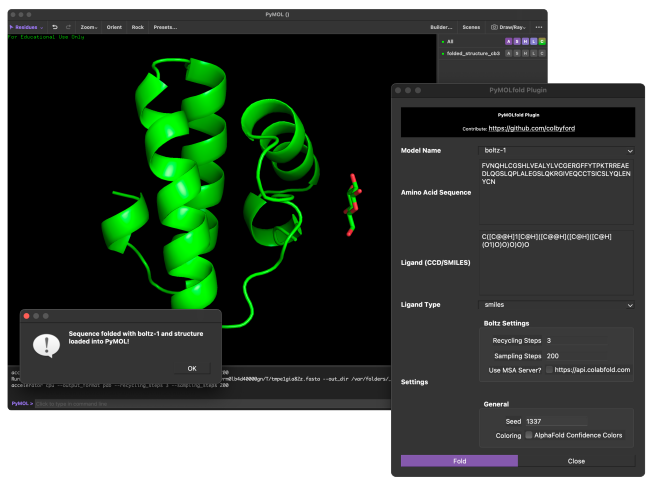

**Fig. 1.** PyMOLfold Plugin

## Introduction

Understanding protein tertiary structures is fundamental to elucidating their functions and interactions. Traditional experimental methods for determining these structures, such as X-ray crystallography and NMR spectroscopy, are often time-consuming and resource-intensive. Recent advancements in computational approaches, particularly those leveraging machine learning, have significantly enhanced the accuracy and accessibility of protein structure predictions.

PyMOL is a widely used molecular visualization system in the scientific community, renowned for its versatility in rendering high-quality 3D images of biological macromolecules. To augment PyMOL's capabilities, we developed PyMOLfold, a plugin that integrates state-of-the-art protein language models—ESM3, Boltz-1, and Chai-1—directly into the PyMOL interface. This integration allows researchers to input amino acid sequences and obtain predicted 3D structures within a single platform, thereby streamlining the workflow from sequence analysis to structural visualization.

By providing an intuitive and efficient tool for protein structure prediction, PyMOLfold aims to accelerate research in structural biology, bioinformatics, and related disciplines. The plugin's open-source nature encourages community-driven development and collaboration, fostering the continuous improvement of its features and predictive capabilities.

## Methods

The front end user interface (UI) of PyMOLfold is built with PyQt5 (1), a set of Python bindings to the Qt application framework (2). The backend logic, written in Python 3, calls the respective libraries for the selected model and loads the folded protein output into the PyMOL session. See Figure 1.

When the user inputs a desired amino acid sequence (and, optionally, a SMILES string for a small molecule), the backend logic prepares the sequence(s) for folding and sends it to the requested model. For example, some model libraries require a file-based input, either a JSON or FASTA file. For those models, PyMOLfold will create temporary a FASTA file (e.g. `>protein|name=chain_A...`) or a JSON file (e.g. `{"proteinChain": {"sequence": ...}`) with the appropriate descriptive information. For other models, PyMOLfold will simply relay the sequence(s) directly to the Python library's folding function.

**Supported Models.** PyMOLfold supports local protein folding using the following models:

- **Boltz-1** - A fully open-source model from the MIT Jameel Clinic that focuses on accurately modeling complex biomolecular interactions (3).

- **Chai-1** - A multi-modal foundation model from Chai Discovery that unifies the prediction of proteins, small molecules, DNA, RNA, and other molecules (4).

- **Protenix** - A trainable PyTorch reproduction of AlphaFold3 from ByteDance (5).

Each of the above models are installed via their respective Python libraries inside the user's PyMOL environment. These models predict protein structures (and complexes with other molecules) using the user's local hardware, including GPUs.

Also, the plugin supports remote model access through the ESM3 application programming interface (API) from EvolutionaryScale Forge (6). This supports multiple model sizes from the ESM3 language model family and retrieves the model weights through an API key.

**Database Retrieval.** While PyMOL already includes the functionality to download experimentally derived structures from the Protein Data Bank (7) using a PDB ID, there are additional databases that house predicted structures that are not included in the Protein Data Bank.

For instance, the AlphaFold Protein Structure Database (AlphaFoldDB), developed by Google DeepMind and EMBL-EBI, houses over 200 million predicted structures (8). Also, ModelArchive is an online repository for computational structure models that are not based on experimental data, often tied to a project or paper (9).

In PyMOLfold, users can retrieve structure from AlphaFoldDB or ModelArchive using accession IDs. For example, a UniProt ID like `Q5VSL9` or a DOI code such as `ma-osf-ppp2r2a`, respectively.

UniProt is the comprehensive database of protein sequences and functional information (10). In PyMOLfold, users may also load in sequence information into the UI using a UniProt search. This allows users to retrieve a protein's sequence, alter any residues of interest, and then have it folded using one of the supported models.

**Installation.** PyMOLfold can be installed directly from the PyMOL interface through the Plugin Manager. Users should either download the .zip file of the latest release of PyMOLfold or copy the URL of it. Then, in the Plugin Manager, the user can install from the downloaded file or paste in the URL to install from the online location. See Figure 2.

Next, the user will need to install the desired folding model libraries into PyMOL's Conda environment. This can be done by locating the Python executable for the PyMOL application and then running `pip install ...` for the desired package.

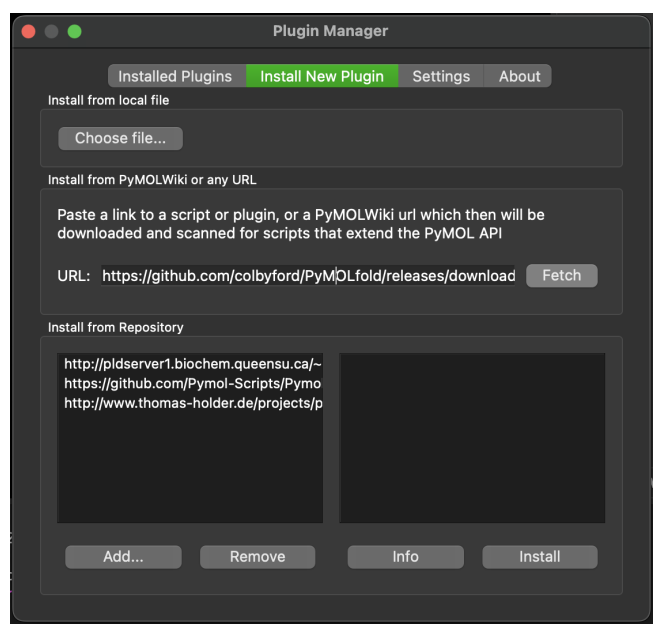

**Fig. 2.** PyMOLfold installation through the Plugin Manager.

For example, `/home/<username>/biotools/pymol/bin/python -m pip install esm` (on Linux).

Also, Conda environment files are provided in the GitHub repository for each of the supported models if users have issues installing in their main version of PyMOL. Note that it may be impossible to install multiple folding libraries in the same Conda environment due to their conflicting dependencies.

More information around the installation process can be found in the GitHub repository mentioned in the Code and Supplementary Materials section.

**Usage.** From the PyMOL user interface, select the Plugin ▶ PyMOLfold from the menu bar. This will open the PyMOLfold UI. Select the desired model with which you wish to fold, and paste in an amino acid sequence (and, optionally, a ligand sequence for some models). Then, set the desired settings and click the *Fold* button. See Figure 3.

In a few seconds, the sequence(s) will be folded and the structure will be loaded into PyMOL. Also, users have the option to color the structure based on AlphaFold-style confidence levels. Once loaded, the structure is fully interactive and customizable as a normal PDB-based structure. Users can then save the molecule locally.

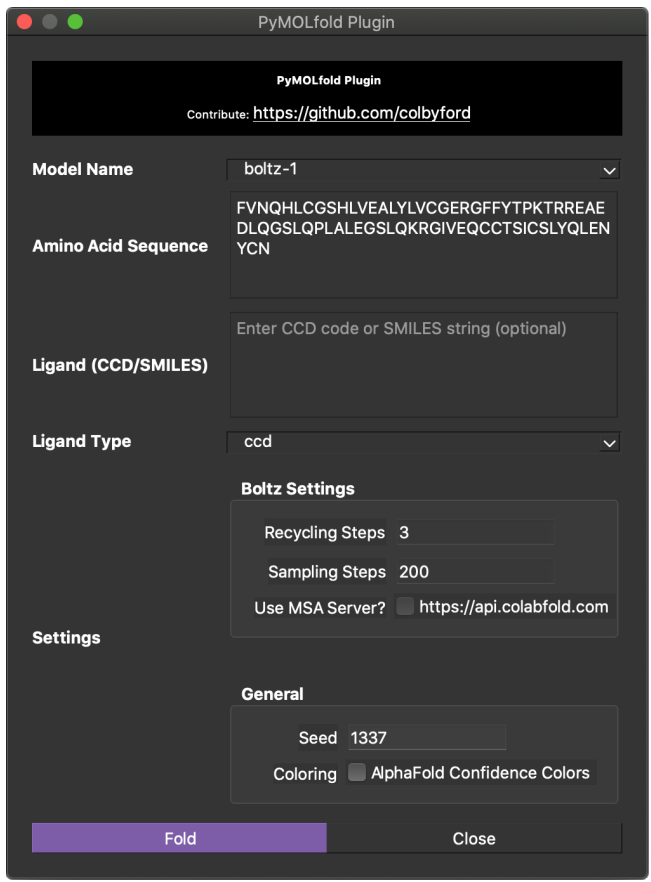

**Fig. 3.** Example usage of the Boltz-1 model.

## Discussion

Previously, protein folding models like AlphaFold were originally quite cumbersome to run as they required a large set of databases and significant software dependencies to perform the sequence alignment process of the algorithm (11, 12). This limited the use of such models to those with the technical acumen to perform the difficult setup and installation and the hardware to run it.

Efforts to make protein folding more accessible, such as ColabFold, have been successful in enabling any user to create structural predictions (13). Also, more recent models, like ESM3, Boltz-1, Chai-1, and others, can be often installed with a single Python command and run with ease.

However, the goal of PyMOLfold is to bridge the gap between having to run structural predictions in one place and then visualize them in another. In other words, a user would need to perform their folding exercise in one environment, then download the .PDB file to a local folder, then open the file in PyMOL to see the result. With PyMOLfold, this is all done in one interface.

**Contributing and Future Work.** PyMOLfold is released as an open-source project under the GPL-3.0 license. We hope that the community will continue to expand its functionality as new models are released.

Upcoming future work includes the support for entering multiple amino acid sequences for running multimer predictions. Also, future capabilities will include the support for running remote predictions through web-based APIs, which will better enable users with lower-grade hardware to get their folding predictions quickly. Also, this will allow users to take advantage of much larger models that would otherwise not fit on consumer-grade GPUs locally.

## Contributors

Author CTF created the original PyMOLfold plugin with support for ESM3. Author US added support for Boltz-1 and Chai-1 and ligand support. CTF and US contributed to the database download functionality. All authors wrote and approved this manuscript.

## Declaration of Interests

Author CTF is the owner of Tuple, LLC, a biotechnology consulting firm. The remaining authors declare that the research was conducted in the absence of any commercial or financial relationships that could be construed as a potential conflict of interest.

## Acknowledgments

We thank Heidong Xianhua from ByteDance for help with implementing the Protenix model. We thank Neil Thomas from EvolutionaryScale for assistance in troubleshooting ESM3 multimer predictions. We also thank the open-source community for their support in testing this software on various operating systems and environments.

## Funding Statement

No external funding was used for this work.

## Code and Supplementary Materials

All code, data, results, and additional analyses are openly available on GitHub at:
https://github.com/colbyford/PyMOLfold.